\newcommand\SCC{H$_8$C$_4$SO$_2\cdot$Cu$_2$Cl$_{4}$}
\newcommand\SCCB{H$_8$C$_4$SO$_2\cdot$Cu$_2$(Cl$_{1-x}$Br$_x$)$_{4}$}

\documentclass[prb,twocolumn,floatfix,preprintnumbers,amsmath,amssymb,superscriptaddress,showpacs]{revtex4}

\usepackage{graphicx}% Include figure files
\usepackage{dcolumn}% Align table columns on decimal point
\usepackage{bm}% bold math
\begin{document}

\title{Disorder instability of the magnon condensate in a frustrated spin ladder.}

\author{E. Wulf}
\affiliation{Neutron Scattering and Magnetism Group, Laboratory for Solid State Physics, ETH Z\"urich, Z\"urich, Switzerland}
\author{S. M\"uhlbauer}
\affiliation{Neutron Scattering and Magnetism Group, Laboratory for Solid State Physics, ETH Z\"urich, Z\"urich, Switzerland}
\author{T. Yankova}
\affiliation{Neutron Scattering and Magnetism Group, Laboratory for Solid State Physics, ETH Z\"urich, Z\"urich, Switzerland}
\affiliation{Permanent address: Chemistry Dept., M. V. Lomonosov
Moscow State University, Moscow, Russia}
\author{A. Zheludev}
\affiliation{Neutron Scattering and Magnetism Group, Laboratory for Solid State Physics, ETH Z\"urich, Z\"urich, Switzerland}

\date{\today}

\begin{abstract}
The effect of disorder is studied on the field-induced quantum phase
transition in the frustrated spin-ladder compound \SCCB\ using bulk
magnetic and thermodynamic measurements. The parent material ($x=0$)
is a quantum spin liquid, which in applied fields is known to form a
magnon condensate with long-range helimagnetic order. We show that
bond randomness introduced by a chemical substitution on the
non-magnetic halogene site destroys this phase transition at very
low concentrations, already for $x=0.01$. The extreme fragility of
the magnon condensate is attributed to {\it random frustration} in
the incommensurate state.
\end{abstract}


\maketitle

\section{Introduction}

Three-dimensional (3D) long-range order is robust and usually highly
resistant to weak spatially random perturbations. In particular,
this applies to field-induced magnetic order in gapped quantum spin
systems. This type of quantum phase transition has recently
attracted a great deal of attention due to its interpretation in
terms of a Bose-Einstein condensation (BEC) of
magnons.\cite{Giamarchi1999,Giamarchi2008} In magnetic materials,
disorder can be introduced by randomizing the strength of magnetic
bonds. In simple models, its primary effect is to produce a random
potential for the condensing magnons. The result is a qualitatively
new phase with non-zero magnetization and susceptibility, but only
short range correlations, the so-called magnetic Bose
glass.\cite{Giamarchi1988,Fisher1989} However, unless the disorder
is very strong, the ordered BEC phase is expected to re-emerge at
higher fields.\cite{Fisher1989,Roscilde2006,Nohadani2005,Yu2010} An
intriguing question is whether {\it weak} randomness in quantum
magnets can go beyond creating a random magnon potential, in a way
that disrupts the formation of the condensate altogether?

\begin{figure*}
\includegraphics[width=0.95\textwidth]{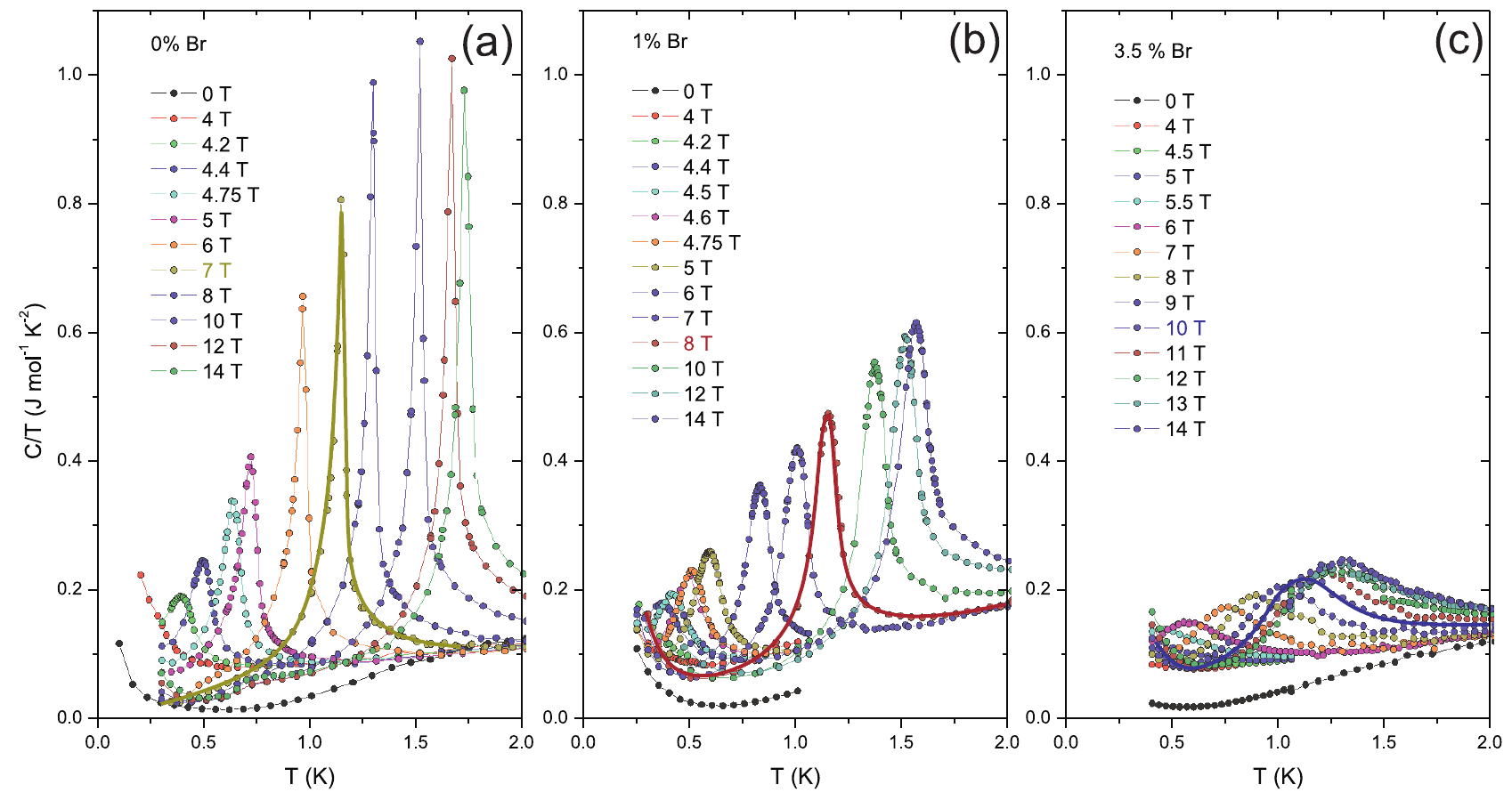}
\caption{(Color online) Connected symbols: specific heat measured as a function
of temperature at various applied fields in SCX with $x=0$ (a), $x=0.01$ (b)
and $x=0.035$ (c). The heavy solid curves are guides to the eye emphasizing
curves peaked at $T\sim 1$~K for easy visual comparison. \label{spec}}
\end{figure*}

In the present work we investigate the effect of bond strength
disorder on the {\it geometrically frustrated} gapped quantum magnet
\SCC\ (SCC for short).\cite{Garlea2008}  Bond disorder is introduced
by a chemical substitution on the non-magnetic site in the
isostructural \SCCB\ (hereafter abbreviated as SCX). We find that
extremely low Br concentrations have a catastrophic effect on the
phase transition and magnetic properties. This behavior is
attributed to {\it frustration disorder}, enhanced by the
incommensurate and quasi-one-dimensional spin correlations in the
parent compound.

To our knowledge, the parent compound SCC is quite unique being a
strongly quasi-one-dimensional gapped Heisenberg spin system where
geometric frustration is strong enough to cause dynamic spin
correlations to be incommensurate. From the point of view of
structural chemistry, it belongs to a large and diverse family of
organic complexes involving transition metal halogenides (see, for
example, Refs.~\onlinecite{Stone2001,Masuda2006,Yuunpublished}).
Antiferromagnetic (AF) interactions between the $S=1/2$ Cu$^{2+}$
ions  are carried by Cu-X-X-Cu superexchange pathways (X=halogen
ion). The resulting spin network can be described as four-leg ``spin
tubes'' with a high degree of geometric
frustration,\cite{Garlea2008,Zheludev2009} and weak inter-tube
coupling. The ground state is a spin liquid, with a gap $\Delta \approx
0.5$~meV in the magnetic excitation spectrum. The exact spin
Hamiltonian is presently unknown. However, due to the strong
one-dimensionality, all relevant low-energy low-temperature
properties are governed by the lowest-energy triplet of
magnons.\cite{Zheludev2008}. in SCC, the latter have a very large
spin velocity $v$ along the $c$ axis, $\hbar vc\approx 14$~meV. A
key observation is that, due to the geometric frustration, the
minimum of dispersion is at an {\it incommensurate} wave vector $q_0
c/(2\pi)=0.48$. In magnetic fields exceeding $H_c=3.75$~T a
condensation of magnons \cite{Fujisawa} produces an incommensurate
phase with spontaneous 3D long-range helimagnetic order of spin
components transverse to the applied fields and a propagation vector
$(0.78,0,0.48)$.\cite{Garlea2009,Zheludev2009} This pitch of the
spin spiral corresponds to the incommensurate  dynamic spin
correlations in zero field.

\section{Experimental procedures and results}

To investigate the effect of disorder on this field-induced phase
transition, we focused on the chemically modified material SCX, with
the Br concentration $x$ ranging 0--5\%. The idea is that
substituting Cl$^-$ by the larger-radius Br$^-$ will affect the
Cu-Cu superexchange coupling locally.\cite{Manaka2001} Since the
relevant Cu-X-X-Cu bond angles are far from 90$^\circ$,
Kanamori-Goodenough rules \cite{KanamoriGoodenough} suggest that the
AF nature of the interaction will be preserved, although its
magnitude may be significantly affected. A random distribution of Br
substitution sites will thus result in a randomization of AF bond
strengths.

A series of well-faceted single crystal samples of typical mass
100~mg were grown by temperature gradient method from ethanol
solution with varying relative Cl/Br content. For each nominal Br
concentration $x$, the samples were characterized by X-ray
diffraction using a Bruker Apex-II single crystal diffractometer and
by micro-elemental chemical analysis (Sch\"oniger method). We found
all crystals to be isostructural to the parent compound. Atomic
positions were refined together with the actual Br concentration,
assuming a uniform distribution of substitution sites. The fitted Br
content is in good agreement with chemical data and the nominal Br
concentration in solution. As far as the lattice parameters are
concerned, the strongest effect of Br substitution is on the $c$
(ladder-leg) axis, which shows a linear increase from 10.035~\AA\ to
10.050~\AA, as $x$ varies from 0 to 5\%.

\subsection*{Characterization of the disorder-free material}

The clearest signature of the field-induced ordering transition in
disorder-free SCC is the lambda-anomaly in the temperature
dependence of specific heat.\cite{Fujisawa} Fig.~\ref{spec}a shows
the heat capacity measured in a 3.5~mg $x=0$ sample as a function of
temperature in several fields applied along the crystallographic $c$
axis, using a Quantum Design PPMS with a dilution refrigerator
insert. The sharp lambda anomaly is quite prominent and allows to
trace the phase boundary on the $H-T$ phase diagram, as shown in
solid symbols in Fig.~\ref{phase}.  The phase transition is also
manifest in the field dependence of magnetic susceptibility
$\chi=dM/dH$ (Fig.~\ref{sus}, solid symbols), measured for $H\| c$
on a 25~mg crystal using a Quantum Design SQUID magnetometer with an
IQuantum $^3$He refrigerator \footnote{From all curves shown in this
figure we have subtracted the contribution of paramagnetic
impurities always present on the sample surface. This contribution
is largest at zero field where it diverges at $T\rightarrow 0$, but
is exponentially negligible at $g\mu_BH\gg \kappa T$ due to
saturation.
It was determined by fitting the magnetization data below $H=1$~T to the $\tanh(g\mu_B J H/\kappa T)$ form, expected for paramagnetic spins.}. The transition appears as a sharp peak, separating the
gapped phase at low fields and the ordered, susceptible high-field
state. For a BEC of magnons, magnetization $M$ is mapped onto Boson
density, and susceptibility $\chi$ - on the compressibility of the
Bose gas.\cite{Giamarchi2008} In these terms, the transition is that
between an incompressible ``Mott-insulator'' magnon phase below
$H_c$, to a compressible Bose condensate of magnons at high fields.
Note that an elevated temperature of 800~mK, the step in the
$\chi(H)$ curve is broadened, but a sharp peak marking the phase
transition persists. The third manifestation of the 3D ordering
transition is found in $M(T)$ curves measured at fixed fields
applied along the $c$ axis (Fig.~\ref{sus}, inset). Upon cooling
through the ordering temperature, the magnetization (Boson density)
shows an almost discontinuous jump following a well-defined minimum,
which is a known characteristic signature of BEC.\cite{Nikuni2000}

\subsection*{Effect of disorder}

The central result of this work is that Br substitution at very low
concentrations {\it destroys} the thermodynamic field-induced
transition described above. Already for $x=1\%$ the lambda anomalies
in the $C(T)$ curves at all fields are replaced by visibly broadened
peaks (Fig.~\ref{spec}b). For $x=3.5\%$ the peak widths increase
further (Fig.~\ref{spec}c). This behavior indicates that the sharp
transition for $x=0$ gives way to a crossover in disordered samples,
presumably to a state with short range order. Indeed, preliminary
neutron diffraction experiments on deuterated samples, performed at
the TASP 3-axis spectrometer at Paul Scherrer Institut and at the
D23 diffractometer at ILL, failed to observe any magnetic
diffraction peaks associated with helimagnetic order. A thorough
search for such reflections was performed in the $(h,0,l)$
reciprocal-space plane for $0.45<l<0.55$, at $T=50$~mK and $H=6$~T.
\footnote{E. Wulf, E. Ressouche, S. M\"uhlbauer, S. Gvasaliya, T.
Yankova and A. Zheludev, unpublished} While it can not be fully
excluded that the peak absence is not due to disorder, but to a
drastic change of the propagation vector along the chain axis, due
to the large spin wave velocity this scenario appears unlikely.

If the high-field phase is indeed short-range ordered, one can
expect diffuse magnetic scattering in the form of anisotropically
broadened peaks at Bragg positions. In SCX no such scattering could
be found, but this result can not be considered conclusive.
Unfortunately, experimental limitation would have made it
unobservable even if it were present. In the disorder-free material
where the peaks are sharp, they are extremely weak and correspond to
only $0.04$~$\mu_\mathrm{B}$ ordered moment.\cite{Garlea2008} Even a
small broadening due to short-range order in SCX would render them
invisible above the rather high intrinsic background.

\begin{figure}
\includegraphics[width=0.95\columnwidth]{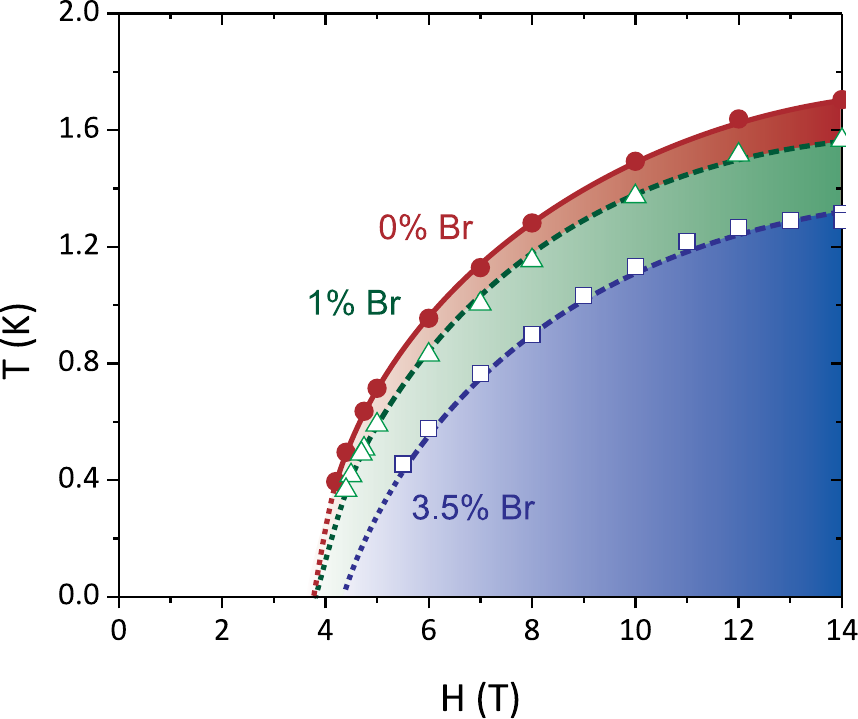}
\caption{(Color online) Solid symbols: $H-T$ phase boundary of SCC
deduced from specific heat data. Open symbols: field dependence of
crossover temperatures in SCX for  $x=0.01$and $x=0.035$. The latter
are defined as positions of broad specific heat maxima in
constant-field scans. Lines are guides for the eye.\label{phase}}
\end{figure}

Returning to the specific heat data in(Fig.~\ref{spec}, we note that
even as the peaks broaden upon Br substitution, their positions are
not subject to a substantial shift. The corresponding crossover
temperatures at each field were identified with the maxima in
$C(T)$, obtained in empirical narrow-range Lorentzian fits. The thus
determined crossover lines, for $x=1\%$ and $x=3.5\%$, are shown in
open symbols in Fig.~\ref{phase}.

\begin{figure}
\includegraphics[width=\columnwidth]{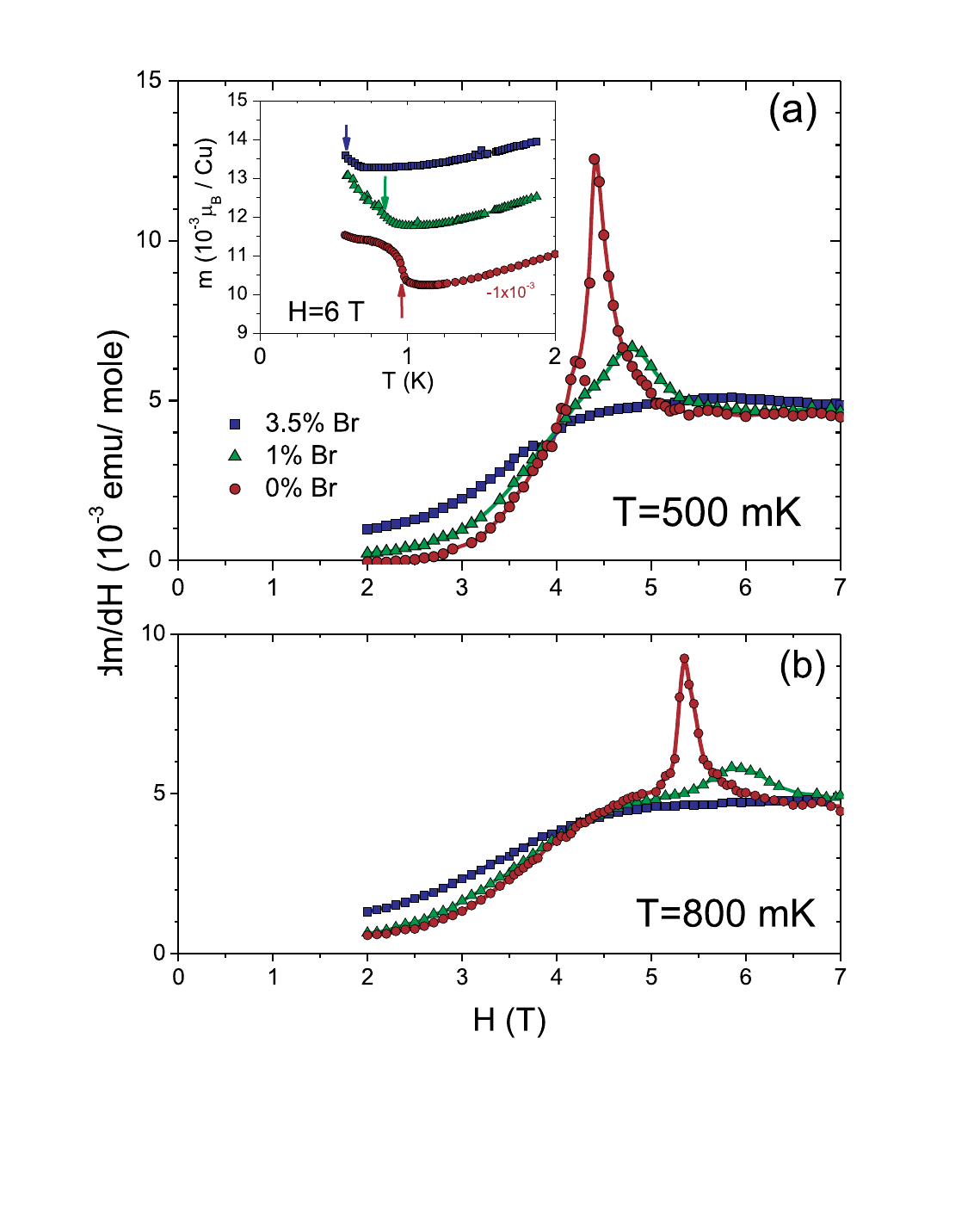}
\caption{(Color online) Field dependence of DC magnetic
susceptibility measured  in SCX with $x=0$, $x=0.01$ and $x=0.035$,
at $T=500$~mK (a) and $T=800$~mK (b). Lines are guides for the eye.
Inset: magnetization as a function of temperature measured a a
constant field $H=6~\mathrm{T}>H_c$. The $x=0$ data set (circles)
has been shifted down by $1\cdot 10^{-3}\mu_\mathrm{B}$/Cu for
visualization purposes. The arrows indicate the positions of $C/T$
maxima for the respective Br concentrations. \label{sus}}
\end{figure}

The effect of chemically introduced randomness on the magnetic
properties is no less dramatic.  The field dependence of magnetic
susceptibility for the disordered samples are shown in
Fig.~\ref{sus}. At all temperatures, the sharp peak seen for $x=0$,
is replaced by a broad feature at $x=1\%$, and altogether absent for
$x=3.5\%$. This behavior is accompanied by an overall broadening of
the step in the $\chi(H)$ curve with increasing $x$. The jump in the
temperature dependence of the magnetization is also smeared out by
disorder (inset in Fig.~\ref{sus}). Thus, magnetometric measurements
confirm that the BEC transition in the parent compound is replaced
by a broad crossover in the disordered systems.

\section{Discussion}

The effect of bond disorder on the field-induced BEC of magnons has
been previously studied experimentally in a number of prototype
materials, including
IPA-Cu(Cl$_{1-x}$Br$_x$)$_3$,\cite{Hong2010PRBRC}
Tl$_{1-x}$K$_x$CuCl$_3$ (Ref.~\onlinecite{Yamada2011}) and most
recently
Ni(Cl$_{1-x}$Br$_x$)$_2\cdot$4SC(NH$_2$)$_2$.\cite{Yuunpublished}
It was found that disorder may considerably affect the phase
boundary, modify the critical behavior and give rise to a
compressible disordered state just below the 3D ordering transition.
However, in {\it all} of these materials, the transition itself is
preserved, even for a large concentration of disorder sites that
significantly alters the critical field and temperature. The
strength and sharpness of anomalies in thermodynamic properties
remains practically unaffected by disorder. The behavior of SCX is
clearly different. Here the condensate appears to be extraordinarily
fragile with respect to randomness, resulting is qualitative
contract between thermodynamic anomalies in the pure and disordered
samples. It is this contrast that we emphasize as the main
experimental finding, since weak or broadened anomalies {\it per se}
do not necessarily signify the absence of a phase transition (see,
for example, Refs.~\onlinecite{Oosawa2001,Nellutla2010}). Below we
propose a qualitative model that attributes this behavior to a
geometric frustration of magnetic interactions and the
incommensurate nature of spin correlations.

\begin{figure}
\includegraphics[width=0.95\columnwidth]{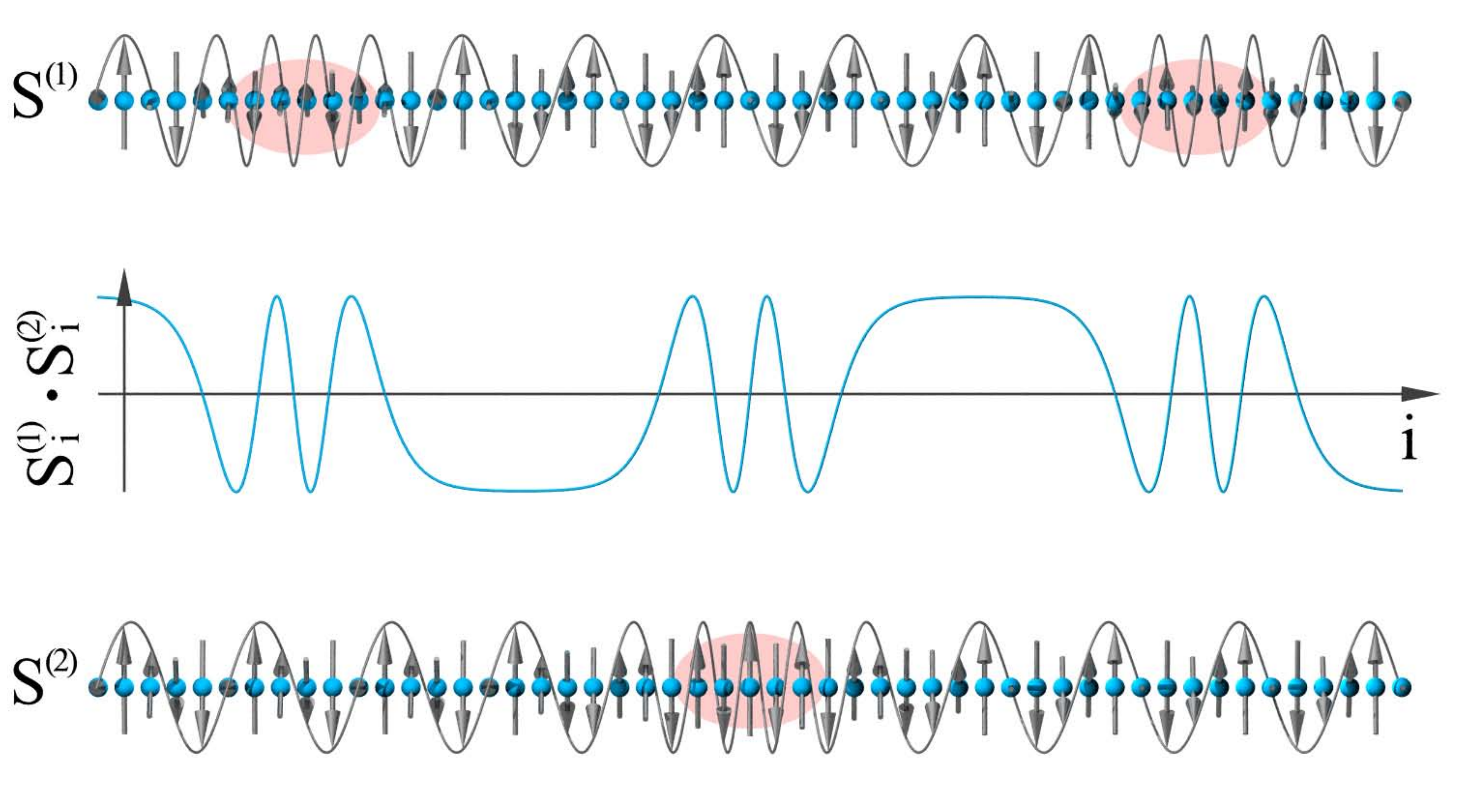}
\caption{A cartoon representation of the random frustration
mechanism. The plot in the center shows the exchange energy (scalar
product) of interactings spins from two adjacent one-dimensional
helimagnets. Defects (shaded ellipses) induce randomly positioned
phase slips, which ensures that inter-chain interactions average to
zero. \label{cartoon}}
\end{figure}

Consider a single geometrically frustrated spin tube in SCC. In a
magnetic field that exceeds $H_c$, it is a gapless Luttinger spin
liquid \cite{Giamarchibook} with a correlation length that diverges
at $T\rightarrow 0$. These increasingly long-range correlations will
be incommensurate and helimagnetic. Their pitch $\phi_0=q_0 c$ will
be defined by the magnitude of geometric frustration of interactions
in the spin tube. A random distribution of Br sites in SCX will
result in a random distribution of the frustration ratios along the
spin tube, and, in turn, a randomized phase of the spiral spin
correlations. For site $j$, we can write $\phi(j)=\phi_0 j
+\sum_{i<j}\delta\phi_i$, where $\delta\phi_i$ are defect-induced
phase slips. Now consider the interaction of two adjacent spin
tubes, assuming no correlations between their substitution sites.
Due to the random phase slips in the spin spirals in each spin tube,
inter-tube exchange interactions will average to zero at the mean
field level. Thus, 3D interactions that are critical for
establishing long-range magnetic order at a finite temperature will
be entirely wiped out by the {\it random frustration} introduced by
bond disorder. This scenario is schematically illustrated in
Fig.~\ref{cartoon}. While it is not directly supported by the
experimental data, it appears very plausible in the particular case
of SCX. Indeed, in our material, the inter-chain interactions are
known to be weak.\cite{Garlea2008} At the same time, due to the very
large spin velocity (temperature equivalent of about 150~K),
one-dimensional correlations are already very well developed at the
experimental temperatures of a few Kelvin. In other words, upon
cooling in a field, individual spin tubes are practically ordered,
and it is the the inter-tube ordering that occurs at the transition
point in the disorder-free material.

It is important to emphasize that in the proposed scenario even {\it
weak} and sparse defects will suppress 3D order. Indeed, even small
phase slips induced by each impurity will, over large distances,
lead to an accumulation of phase difference between adjacent spin
tubes. This is a specific consequence of the incommensurate and
quasi-1D nature of the disorder-free system.

Incommensurability and frustration being the key ingredients of our
model makes it {\it distinct} from the basic mechanism of Bose
Glasses. The latter emerges from the mapping the magnetic system
onto a lattice gas of magnons with a spatially random chemical
potential. Neither incommensurability or frustration are explicitly
present in the resulting bosonic Hamiltonian, and are therefore
overlooked by the approach. For our newly proposed mechanism, there
are numerous theoretical questions to be further addressed. Is there
a critical strength of frustration disorder that is required to
disrupt the magnon condensate? What are the critical properties of
the order/disorder transition induced by random frustration? How
does the random frustration physics interplay with that of the
``conventional'' Bose glass?

\acknowledgements

This work is partially supported by the Swiss National Fund under
project 2-77060-11 and through Project 6 of MANEP. We thank Dr. V.
Glazkov (Kapitza Institute, Russian Acad. Sci.) for his involvement
at the early stages of this project.

%\bibliography{C:/home/zhelud/bib/azbib}

\begin{thebibliography}{23}
\expandafter\ifx\csname
natexlab\endcsname\relax\def\natexlab#1{#1}\fi
\expandafter\ifx\csname bibnamefont\endcsname\relax
  \def\bibnamefont#1{#1}\fi
\expandafter\ifx\csname bibfnamefont\endcsname\relax
  \def\bibfnamefont#1{#1}\fi
\expandafter\ifx\csname citenamefont\endcsname\relax
  \def\citenamefont#1{#1}\fi
\expandafter\ifx\csname url\endcsname\relax
  \def\url#1{\texttt{#1}}\fi
\expandafter\ifx\csname urlprefix\endcsname\relax\def\urlprefix{URL
}\fi \providecommand{\bibinfo}[2]{#2}
\providecommand{\eprint}[2][]{\url{#2}}

\bibitem[{\citenamefont{Giamarchi and Tsvelik}(1999)}]{Giamarchi1999}
\bibinfo{author}{\bibfnamefont{T.}~\bibnamefont{Giamarchi}} \bibnamefont{and}
  \bibinfo{author}{\bibfnamefont{A.~M.} \bibnamefont{Tsvelik}},
  \bibinfo{journal}{Phys. Rev. B} \textbf{\bibinfo{volume}{59}},
  \bibinfo{pages}{11398} (\bibinfo{year}{1999}).

\bibitem[{\citenamefont{Giamarchi et~al.}(2008)\citenamefont{Giamarchi, R\"uegg,
  and Tchernyshyov}}]{Giamarchi2008}
\bibinfo{author}{\bibfnamefont{T.}~\bibnamefont{Giamarchi}},
  \bibinfo{author}{\bibfnamefont{C.}~\bibnamefont{R\"uegg}}, \bibnamefont{and}
  \bibinfo{author}{\bibfnamefont{O.}~\bibnamefont{Tchernyshyov}},
  \bibinfo{journal}{Nature Physics} \textbf{\bibinfo{volume}{4}},
  \bibinfo{pages}{198} (\bibinfo{year}{2008}).

\bibitem[{\citenamefont{Giamarchi and Schulz}(1988)}]{Giamarchi1988}
\bibinfo{author}{\bibfnamefont{T.}~\bibnamefont{Giamarchi}} \bibnamefont{and}
  \bibinfo{author}{\bibfnamefont{H.~J.} \bibnamefont{Schulz}},
  \bibinfo{journal}{Phys. Rev. B} \textbf{\bibinfo{volume}{37}},
  \bibinfo{pages}{325} (\bibinfo{year}{1988}).

\bibitem[{\citenamefont{Fisher et~al.}(1989)\citenamefont{Fisher, Weichman,
  Grinstein, and Fisher}}]{Fisher1989}
\bibinfo{author}{\bibfnamefont{M.~P.~A.} \bibnamefont{Fisher}},
  \bibinfo{author}{\bibfnamefont{P.~B.} \bibnamefont{Weichman}},
  \bibinfo{author}{\bibfnamefont{G.}~\bibnamefont{Grinstein}},
  \bibnamefont{and} \bibinfo{author}{\bibfnamefont{D.~S.}
  \bibnamefont{Fisher}}, \bibinfo{journal}{Phys. Rev. B}
  \textbf{\bibinfo{volume}{40}}, \bibinfo{pages}{546} (\bibinfo{year}{1989}).

\bibitem[{\citenamefont{Roscilde}(2006)}]{Roscilde2006}
\bibinfo{author}{\bibfnamefont{T.}~\bibnamefont{Roscilde}},
  \bibinfo{journal}{Phys. Rev. B} \textbf{\bibinfo{volume}{74}},
  \bibinfo{pages}{144418} (\bibinfo{year}{2006}).

\bibitem[{\citenamefont{Nohadani et~al.}(2005)\citenamefont{Nohadani, Wessel,
  and Haas}}]{Nohadani2005}
\bibinfo{author}{\bibfnamefont{O.}~\bibnamefont{Nohadani}},
  \bibinfo{author}{\bibfnamefont{S.}~\bibnamefont{Wessel}}, \bibnamefont{and}
  \bibinfo{author}{\bibfnamefont{S.}~\bibnamefont{Haas}},
  \bibinfo{journal}{Phys. Rev. Lett.} \textbf{\bibinfo{volume}{95}},
  \bibinfo{pages}{227201} (\bibinfo{year}{2005}).

\bibitem[{\citenamefont{Yu et~al.}(2010)\citenamefont{Yu, Haas, and
  Roscilde}}]{Yu2010}
\bibinfo{author}{\bibfnamefont{R.}~\bibnamefont{Yu}},
  \bibinfo{author}{\bibfnamefont{S.}~\bibnamefont{Haas}}, \bibnamefont{and}
  \bibinfo{author}{\bibfnamefont{T.}~\bibnamefont{Roscilde}},
  \bibinfo{journal}{Europhys. Lett.} \textbf{\bibinfo{volume}{89}},
  \bibinfo{pages}{10009} (\bibinfo{year}{2010}).

\bibitem[{\citenamefont{Garlea et~al.}(2008)\citenamefont{Garlea, Zheludev,
  Regnault, Chung, Qiu, Boehm, Habicht, and Meissner}}]{Garlea2008}
\bibinfo{author}{\bibfnamefont{V.~O.} \bibnamefont{Garlea}},
  \bibinfo{author}{\bibfnamefont{A.}~\bibnamefont{Zheludev}},
  \bibinfo{author}{\bibfnamefont{L.-P.} \bibnamefont{Regnault}},
  \bibinfo{author}{\bibfnamefont{J.-H.} \bibnamefont{Chung}},
  \bibinfo{author}{\bibfnamefont{Y.}~\bibnamefont{Qiu}},
  \bibinfo{author}{\bibfnamefont{M.}~\bibnamefont{Boehm}},
  \bibinfo{author}{\bibfnamefont{K.}~\bibnamefont{Habicht}}, \bibnamefont{and}
  \bibinfo{author}{\bibfnamefont{M.}~\bibnamefont{Meissner}},
  \bibinfo{journal}{Physical Review Letters} \textbf{\bibinfo{volume}{100}},
  \bibinfo{pages}{037206} (\bibinfo{year}{2008}).

\bibitem[{\citenamefont{Stone et~al.}(2001)\citenamefont{Stone, Zaliznyak,
  Reich, and Broholm}}]{Stone2001}
\bibinfo{author}{\bibfnamefont{M.~B.} \bibnamefont{Stone}},
  \bibinfo{author}{\bibfnamefont{I.}~\bibnamefont{Zaliznyak}},
  \bibinfo{author}{\bibfnamefont{D.~H.} \bibnamefont{Reich}}, \bibnamefont{and}
  \bibinfo{author}{\bibfnamefont{C.}~\bibnamefont{Broholm}},
  \bibinfo{journal}{Phys. Rev. B} \textbf{\bibinfo{volume}{64}},
  \bibinfo{pages}{144405} (\bibinfo{year}{2001}).

\bibitem[{\citenamefont{Masuda et~al.}(2006)\citenamefont{Masuda, Zheludev,
  Manaka, Regnault, Chung, and Qiu}}]{Masuda2006}
\bibinfo{author}{\bibfnamefont{T.}~\bibnamefont{Masuda}},
  \bibinfo{author}{\bibfnamefont{A.}~\bibnamefont{Zheludev}},
  \bibinfo{author}{\bibfnamefont{H.}~\bibnamefont{Manaka}},
  \bibinfo{author}{\bibfnamefont{L.-P.} \bibnamefont{Regnault}},
  \bibinfo{author}{\bibfnamefont{J.-H.} \bibnamefont{Chung}}, \bibnamefont{and}
  \bibinfo{author}{\bibfnamefont{Y.}~\bibnamefont{Qiu}},
  \bibinfo{journal}{Phys. Rev. Lett.} \textbf{\bibinfo{volume}{96}},
  \bibinfo{pages}{047210} (\bibinfo{year}{2006}).

\bibitem[{Yin()}]{Yuunpublished}
\bibinfo{note}{R. Yu, L. Yin, N.~S. Sullivan, J.~S. Xia, C. Huan, A. Paduan-Filho, N.~F. Oliveira Jr., St. Haas, A. Steppke, C.~F. Miclea, F. Weickert, R. Movshovich, E.-D. Mun, V.~S. Zapf, and T. Roscilde, arXiv:1109.4403v2}.

\bibitem[{\citenamefont{Zheludev et~al.}(2009)\citenamefont{Zheludev, Garlea,
  Tsvelik, Regnault, Habicht, Kiefer, and Roessli}}]{Zheludev2009}
\bibinfo{author}{\bibfnamefont{A.}~\bibnamefont{Zheludev}},
  \bibinfo{author}{\bibfnamefont{V.~O.} \bibnamefont{Garlea}},
  \bibinfo{author}{\bibfnamefont{A.}~\bibnamefont{Tsvelik}},
  \bibinfo{author}{\bibfnamefont{L.-P.} \bibnamefont{Regnault}},
  \bibinfo{author}{\bibfnamefont{K.}~\bibnamefont{Habicht}},
  \bibinfo{author}{\bibfnamefont{K.}~\bibnamefont{Kiefer}}, \bibnamefont{and}
  \bibinfo{author}{\bibfnamefont{B.}~\bibnamefont{Roessli}},
  \bibinfo{journal}{Phys. Rev. B} \textbf{\bibinfo{volume}{80}},
  \bibinfo{pages}{214413} (\bibinfo{year}{2009}).

\bibitem[{\citenamefont{Zheludev et~al.}(2008)\citenamefont{Zheludev, Garlea,
  Regnault, Manaka, Tsvelik, and Chung}}]{Zheludev2008}
\bibinfo{author}{\bibfnamefont{A.}~\bibnamefont{Zheludev}},
  \bibinfo{author}{\bibfnamefont{V.~O.} \bibnamefont{Garlea}},
  \bibinfo{author}{\bibfnamefont{L.-P.} \bibnamefont{Regnault}},
  \bibinfo{author}{\bibfnamefont{H.}~\bibnamefont{Manaka}},
  \bibinfo{author}{\bibfnamefont{A.}~\bibnamefont{Tsvelik}}, \bibnamefont{and}
  \bibinfo{author}{\bibfnamefont{J.-H.} \bibnamefont{Chung}},
  \bibinfo{journal}{Phys. Rev. Lett.} \textbf{\bibinfo{volume}{100}},
  \bibinfo{pages}{157204} (\bibinfo{year}{2008}).

\bibitem[{\citenamefont{Fujisawa}(2006)}]{Fujisawa}
\bibinfo{author}{\bibfnamefont{M.}~\bibnamefont{Fujisawa}}, Ph.D. thesis,
  \bibinfo{school}{Tokyo Institute of Technology} (\bibinfo{year}{2006}).

\bibitem[{\citenamefont{Garlea et~al.}(2009)\citenamefont{Garlea, Zheludev,
  Habicht, Meissner, Grenier, Regnault, and Ressouche}}]{Garlea2009}
\bibinfo{author}{\bibfnamefont{V.~O.} \bibnamefont{Garlea}},
  \bibinfo{author}{\bibfnamefont{A.}~\bibnamefont{Zheludev}},
  \bibinfo{author}{\bibfnamefont{K.}~\bibnamefont{Habicht}},
  \bibinfo{author}{\bibfnamefont{M.}~\bibnamefont{Meissner}},
  \bibinfo{author}{\bibfnamefont{B.}~\bibnamefont{Grenier}},
  \bibinfo{author}{\bibfnamefont{L.-P.} \bibnamefont{Regnault}},
  \bibnamefont{and}
  \bibinfo{author}{\bibfnamefont{E.}~\bibnamefont{Ressouche}},
  \bibinfo{journal}{Physical Review B} \textbf{\bibinfo{volume}{79}},
  \bibinfo{pages}{060404} (\bibinfo{year}{2009}).

\bibitem[{\citenamefont{Manaka et~al.}(2001)\citenamefont{Manaka, Yamada,
  Hagiwara, and Tokunaga}}]{Manaka2001}
\bibinfo{author}{\bibfnamefont{H.}~\bibnamefont{Manaka}},
  \bibinfo{author}{\bibfnamefont{I.}~\bibnamefont{Yamada}},
  \bibinfo{author}{\bibfnamefont{M.}~\bibnamefont{Hagiwara}}, \bibnamefont{and}
  \bibinfo{author}{\bibfnamefont{M.}~\bibnamefont{Tokunaga}},
  \bibinfo{journal}{Phys. Rev. B} \textbf{\bibinfo{volume}{63}},
  \bibinfo{pages}{144428} (\bibinfo{year}{2001}).

\bibitem[{Kan()}]{KanamoriGoodenough}
\bibinfo{note}{J. B. Goodenough, J. Phys. Chem. Solids {\bf 6}, 287 (1958); J.
  Kanamori, J. Phys. Chem. Solids {\bf 10}, 87 (1959).}

\bibitem[{\citenamefont{Nikuni et~al.}(2000)\citenamefont{Nikuni, Oshikawa,
  Oosawa, and Tanaka}}]{Nikuni2000}
\bibinfo{author}{\bibfnamefont{T.}~\bibnamefont{Nikuni}},
  \bibinfo{author}{\bibfnamefont{M.}~\bibnamefont{Oshikawa}},
  \bibinfo{author}{\bibfnamefont{A.}~\bibnamefont{Oosawa}}, \bibnamefont{and}
  \bibinfo{author}{\bibfnamefont{H.}~\bibnamefont{Tanaka}},
  \bibinfo{journal}{Phys. Rev. Lett.} \textbf{\bibinfo{volume}{84}},
  \bibinfo{pages}{5868} (\bibinfo{year}{2000}).

\bibitem[{\citenamefont{Hong et~al.}(2010)\citenamefont{Hong, Zheludev, Manaka,
  and Regnault}}]{Hong2010PRBRC}
\bibinfo{author}{\bibfnamefont{T.}~\bibnamefont{Hong}},
  \bibinfo{author}{\bibfnamefont{A.}~\bibnamefont{Zheludev}},
  \bibinfo{author}{\bibfnamefont{H.}~\bibnamefont{Manaka}}, \bibnamefont{and}
  \bibinfo{author}{\bibfnamefont{L.-P.} \bibnamefont{Regnault}},
  \bibinfo{journal}{Phys. Rev. B} \textbf{\bibinfo{volume}{81}},
  \bibinfo{pages}{060410} (\bibinfo{year}{2010}).

\bibitem[{\citenamefont{Yamada et~al.}(2011)\citenamefont{Yamada, Tanaka, Ono,
  and Nojiri}}]{Yamada2011}
\bibinfo{author}{\bibfnamefont{F.}~\bibnamefont{Yamada}},
  \bibinfo{author}{\bibfnamefont{H.}~\bibnamefont{Tanaka}},
  \bibinfo{author}{\bibfnamefont{T.}~\bibnamefont{Ono}}, \bibnamefont{and}
  \bibinfo{author}{\bibfnamefont{H.}~\bibnamefont{Nojiri}},
  \bibinfo{journal}{Phys. Rev. B} \textbf{\bibinfo{volume}{83}},
  \bibinfo{pages}{020409} (\bibinfo{year}{2011}). See also comment by A.
  Zheludev and D. H\"uvonen {\it ibid.},  216401 (2011), and the
  reply by Yamada {\it et al.},  {\it ibid.},  216402 (2011).

\bibitem[{\citenamefont{Oosawa et~al.}(2001)\citenamefont{Oosawa, Aruga~Katori, and
  Tanaka}}]{Oosawa2001}
\bibinfo{author}{\bibfnamefont{A.}~\bibnamefont{Oosawa}},
  \bibinfo{author}{\bibfnamefont{H.} \bibnamefont{Aruga~Katori}},
  \bibnamefont{and} \bibinfo{author}{\bibfnamefont{H.}~\bibnamefont{Tanaka}},
  \bibinfo{journal}{Phys. Rev. B} \textbf{\bibinfo{volume}{63}},
  \bibinfo{pages}{134416} (\bibinfo{year}{2001}).

\bibitem[{\citenamefont{Nellutla et~al.}(2010)\citenamefont{Nellutla, Pati, Jo,
  Zhou, Moon, Pajerowski, Yoshida, Janik, Balicas, Lee et~al.}}]{Nellutla2010}
\bibinfo{author}{\bibfnamefont{S.}~\bibnamefont{Nellutla}},
  \bibinfo{author}{\bibfnamefont{M.}~\bibnamefont{Pati}},
  \bibinfo{author}{\bibfnamefont{Y.-J.} \bibnamefont{Jo}},
  \bibinfo{author}{\bibfnamefont{H.~D.} \bibnamefont{Zhou}},
  \bibinfo{author}{\bibfnamefont{B.~H.} \bibnamefont{Moon}},
  \bibinfo{author}{\bibfnamefont{D.~M.} \bibnamefont{Pajerowski}},
  \bibinfo{author}{\bibfnamefont{Y.}~\bibnamefont{Yoshida}},
  \bibinfo{author}{\bibfnamefont{J.~A.} \bibnamefont{Janik}},
  \bibinfo{author}{\bibfnamefont{L.}~\bibnamefont{Balicas}},
  \bibinfo{author}{\bibfnamefont{Y.}~\bibnamefont{Lee}}, \bibnamefont{et~al.},
  \bibinfo{journal}{Phys. Rev. B} \textbf{\bibinfo{volume}{81}},
  \bibinfo{pages}{064431} (\bibinfo{year}{2010}).

\bibitem[{\citenamefont{Giamarchi}(2003)}]{Giamarchibook}
\bibinfo{author}{\bibfnamefont{T.}~\bibnamefont{Giamarchi}},
  \emph{\bibinfo{title}{Quantum Physics in One Dimension}}
  (\bibinfo{publisher}{Clarendon Press}, \bibinfo{year}{2003}).

\end{thebibliography}

\end{document}